\documentclass[12p]{iopart}

\usepackage{graphicx}
\usepackage{multirow, array}
\usepackage{color}
\usepackage{url}
\begin{document}

\title{Convergence and efficiency of angular momentum projection} 

\author{Calvin W. Johnson}
\ead{cjohnson@sdsu.edu}
\author{Changfeng Jiao}
\address{Department of Physics, San Diego State University,
5500 Campanile Drive, San Diego, CA 92182-1233}


\begin{abstract}
In many so-called ``beyond-mean-field'' many-body methods, one creates symmetry-breaking states and then 
projects out states with good quantum number(s); the most important example is angular 
momentum. Motivated by the computational intensity of symmetry restoration, we investigate 
the numerical convergence of two competing methods for angular momentum projection with rotations over Euler angles, the textbook-standard projection through quadrature, 
and a recently introduced projection through linear algebra. We find well-defined patterns of convergence with increasing 
number of mesh points (for quadrature) and cut-offs (for linear algebra).  Because the method of projection through 
linear algebra requires inverting matrices generated on a mesh of Euler angles, we discuss two methods for robustly 
reducing the number of required evaluations. 
{{
Reviewing the literature, we find our inversion involving rotations about the $z$-axis is equivalent to  trapezoidal
`quadrature' commonly used as well as Fomenko projection used for particle-number projection.
The efficiency depends upon the number of angular momentum $J$ to be projected, but in general 
inversion methods, including Fomenko projection/trapezoidal `quadrature' dramatically improve the efficiency.}}
\end{abstract}


\submitto{\jpg}
\maketitle


The general quantum many-body problem is numerically challenging, and a wide-ranging portfolio of methods 
have been developed to tackle it \cite{ring2004nuclear}.   Many, though not all, of these begin with a mean-field or independent-particle 
picture, including quasi-particle methods, in large part because products of single-particle states are conceptually straightforward. 
As we restrict ourselves to systems with a finite and well-defined number 
of fermions, such as nuclei and atoms, we naturally come to antisymmetrized products of single-particle states, 
or Slater determinants and, using second quantization, their occupation-number representations \cite{ring2004nuclear,suhonen2007nucleons}

While symmetries for isolated many-body systems, such as nuclei and atoms, often dictate  exact conservation laws, such as a fixed number of particles and 
conservation of angular momentum,  it can be paradoxically advantageous to disregard these symmetries in the 
mean-field and  restore them later via projection  \cite{ring2004nuclear,bohr1998nuclear2}.  
Examples of these so-called ``beyond mean-field'' methods include 
 projected Hartree-Fock \cite{PhysRev.156.1087} including variation after projection \cite{PhysRevC.71.044313},
 and Hartree-Fock-Bogoliubov \cite{hara1982exact,PhysRevC.59.135,sheikh2000symmetry,Borrajo2016} and projected relativistic mean-field calculations 
 \cite{PhysRevC.79.044312};
the Monte Carlo Shell Model \cite{PhysRevLett.77.3315,abe2013recent}; the projected shell model \cite{hara1995projected,sun1997fortran,PhysRevLett.82.3968}; the
projected configuration-interaction \cite{PhysRevC.79.014311} and related methods \cite{schmid2004use}; and 
projected generator coordinate 
\cite{PhysRevC.65.024304,PhysRevC.74.064309,PhysRevC.78.024309,PhysRevC.81.044311,PhysRevC.81.064323,PhysRevC.83.014308,borrajo2015symmetry,Rodríguez2016}. 

While some of the above methods also break and then restore particle number and/or isospin, in this paper we will focus exclusively upon projecting good angular momentum.  The standard method for projecting out good angular 
momentum is a three-dimensional numerical integral (quadrature) over the Euler angles $(\alpha, \beta, \gamma) = \Omega$.  This requires evaluation of expensive 
matrix elements at many values of $\Omega $. 
(There are also methods which apply polynomials in the angular momentum operators $\hat{J}^2$ , $\hat{J}_z$, and 
$\hat{J}_\pm$, see \cite{RevModPhys.36.966} and 
related papers, e.g. \cite{PhysRevLett.27.439,PhysRevC.88.034305}.  Methods involving rotations are nonetheless commonly used.)


In this paper we continue 
{{
the work of Ref.~\cite{PhysRevC.96.064304}, which showed how one could instead view 
projection as a simple problem in linear algebra involving rotated states. It also  pointed out that the main computational burden 
is in evaluating the Hamiltonian kernel at different Euler angles.  The time of a calculation, and therefore its efficiency,
is directly proportional to the number of evaluations needed to reach a given accuracy.  In this paper we
investigate and compare the efficiency of
}}
   angular momentum projection by both quadrature and by linear algebra.
 Quadrature angular momentum 
projection routinely uses discrete symmetries to reduce the number of evaluations 
\cite{hara1995projected,sun1997fortran,PhysRevC.78.024309,PhysRevC.81.064323},
 and we show how to adopt certain discrete symmetries into linear algebra projection. 
 We also discuss further development of a need-to-know 
approach in linear algebra projection to further reduce the required evaluations. Finally, for both methods  
we investigate the accuracy with respect to the number of evaluations taken. A good test for accuracy is 
the fraction of the wave function for a given angular momentum value $J$, which takes the form of a trace over the norm kernel, which is cheaper to evaluate than  the Hamiltonian kernel.

\section{Two methods for angular momentum projection}

Projection via quadrature and via linear algebra both 
 start with  the rotation operator over the Euler angles
\begin{equation}
\hat{R}(\alpha, \beta,\gamma) = \exp \left( i \gamma \hat{J}_z \right ) \exp \left( i \beta \hat{J}_y \right )  \exp \left( i \alpha \hat{J}_z \right ), \label{rotation}
\end{equation}
with $\hat{J}_z$ and $\hat{J}_y$  the generators of rotations about the $z$ and $y$-axes, respectively. 
The matrix elements of rotation between states of good angular momentum $ | J, M \rangle $ are the Wigner $D$-matrices:
\begin{equation}
{\cal D}^{(J)}_{M,K} (\alpha, \beta,\gamma)
= \langle J, M | \hat{R}(\alpha, \beta,\gamma) | J, K \rangle = 
e^{i \alpha M} d^J_{MK}(\beta) e^{i \gamma K},
\label{WignerD}
\end{equation}
where $d^J_{MK}(\beta)$ is the Wigner little-$d$ function. 
Because the Wigner $D$-matrices form a complete, orthogonal set \cite{edmonds1996angular}, the 
standard method is to use numerical quadrature   to project out states of good angular momentum \cite{ring2004nuclear}. In particular, one generates the overlap or norm 
matrix,
\begin{equation}
N^J_{MK} =   \frac{2J+1}{8 \pi^2} \int d\Omega \, {\cal D}^{(J)*}_{M,K}(\Omega) \left \langle \Psi \left | \hat{R}(\Omega) \right | \Psi \right \rangle \label{normdefnquad} 
\end{equation}
as well as the Hamiltonian
\begin{equation}
H^J_{MK} =   \frac{2J+1}{8 \pi^2} \int d\Omega \, {\cal D}^{(J)*}_{M,K}(\Omega) \left \langle \Psi \left | \hat{H} \hat{R}(\Omega) \right | \Psi \right \rangle \label{hamdefn} 
\end{equation}
where $\hat{H}$ is the many-body Hamiltonian. 
One then solves for each $J$ the generalized eigenvalue problem, with solutions labeled by $r$.
\begin{equation}
\sum_K H^J_{MK}g^{(J)}_{K,r} = E_r \sum_K N^J_{M,K}g^{(J)}_{K,r}, \label{eigen}
\end{equation}
In these calculations, the norm kernel, which is just  the matrix element of the rotation operator 
 $\langle \Psi | \hat{R}(\Omega) | \Psi \rangle$, is significantly cheaper to compute than the Hamiltonian kernel $ \langle \Psi | \hat{H} \hat{R}(\Omega) | \Psi \rangle $, especially as the model space increases in size \cite{PhysRevC.96.064304}. 

There is however another way to project \cite{PhysRevC.96.064304}. 
Before  the integrals over the Euler angles in Eq.~(\ref{normdefnquad},{\ref{hamdefn}) are evaluated,
 notice that
\begin{equation}
\langle \Psi | \hat{R}(\Omega) | \Psi \rangle = \sum_{J,K,M}  {\cal D}^{(J)}_{M,K} (\Omega)N^J_{MK}, \label{linrelnorm} 
\end{equation}
\begin{equation}
\langle \Psi | \hat{H} \hat{R}(\Omega) | \Psi \rangle = \sum_{J,K,M}  {\cal D}^{(J)}_{M,K} (\Omega) H^J_{MK}. \label{linrelham} 
\end{equation}
In other words, the norm kernel
 $\langle \Psi | \hat{R}(\Omega) | \Psi \rangle$ is a linear combination of the norm matrix elements $N^J_{MK}$, and 
the same for  the Hamiltonian kernel relative to the Hamiltonian matrix elements $H^J_{MK}$.  
So instead of using 
 orthogonality of the ${\cal D}$-matrices, one  
 solves Eqn.~(\ref{linrelnorm}) and (\ref{linrelham}) as a linear algebra problem. 
That is, if we label a particular choice of Euler angles $\Omega$ by $i$ and the angular momentum quantum numbers $J,M,K$ by $a$, 
and define 
\begin{eqnarray}
n_i \equiv \langle \Psi | \hat{R}(\Omega_i) | \Psi \rangle , \nonumber\\
D_{ia} \equiv  {\cal D}^{(J_a)}_{M_a,K_a} (\Omega_i), \\
N_a \equiv N^{J_a}_{M_a K_a}, \nonumber 
\end{eqnarray}
we can rewrite Eq.~(\ref{linrelnorm}) simply as 
\begin{equation}
n_i = \sum_a D_{ia} N_a \label{compactLA}
\end{equation}
which can be easily solved for $N_a = N^J_{M,K}$, as long as $D_{ia}$ is invertible, with a similar rewriting of Eq.~(\ref{linrelham}) and 
solution for $H^J_{M,K}$.   The question of invertibility is not a trivial one, and is an important issue in this paper.

A key idea is that the sums (\ref{linrelnorm}), (\ref{linrelham}) are finite.   To justify this, we introduce 
the  fractional `occupation' of the wave function with angular momentum 
$J$, which is the trace of the  fixed-$J$ norm matrix:
\begin{equation}
f_J = \sum_{M} N^J_{M,M}. \label{def_fJ}
\end{equation}
Assuming the original state is normalized, one trivially has 
\begin{equation}
\sum_J f_J = 1.
\label{sumrule}
\end{equation}  

The fractional occupation $f_J$ and its sum rule (\ref{sumrule}) have multiple uses. First, the sum rule is an important check on any calculation.   
Second, as discussed below, $f_J$ acts as an inexpensive measure of convergence with, for example, the 
quadrature mesh, allowing one to find a `right-sized' mesh. 
Finally, one can use the exhaustion of the sum rule to determine a maximum angular 
momentum, $J_\mathrm{max}$, 
in our expansions; in our trials we found Eq.~(\ref{normdefnquad},{\ref{hamdefn})and  (\ref{sumrule})  dominated by a finite and relatively small number of terms,
 far fewer terms than are allowed even in finite model spaces.  As discussed in the next section, we found that fractional occupations below 
 $0.001$ could be safely ignored.
 
 
 {{

\section{Projection by quadrature} 
\label{quadrature}

Because evaluations of the Hamiltonian kernel can be expensive, it is natural to turn to 
optimized quadrature methods for evaluating integrals on the Euler angles $\alpha, \beta, \gamma$. 
Throughout the literature the method of choice for $\beta$-integrals (rotations about the $y$-axis)  is Gauss-Legendre quadrature. One could 
choose Gauss-Jacobi quadrature, as the Wigner little-$d$ function in $\beta$ can be written in terms of Jacobi polynomials \cite{edmonds1996angular}, but the mesh of $\beta_i$ 
depends on specific values of $M,K$, and one would lose any increase in quadrature efficiency through multiple repetitions.

For evaluating integrals over $\alpha$ and $\gamma$ (rotations about the $z$-axis) for triaxial projection,  both Gauss-Legendre  quadrature 
\cite{PhysRevC.79.044312, PhysRevC.81.044311, PhysRevC.83.014308,PhysRevC.81.064323,Rodríguez2016}, 
and trapezoidal  quadrature \cite{hara1995projected,PhysRevC.78.024309} have been used.
 At first glance the latter seems odd, as trapezoidal quadrature is usually a less accurate numerical quadrature 
method.  If we write out the trapezoidal rule, however,
\begin{equation}
         \frac{1}{2\pi}\int^{2\pi}_0 d\gamma e^{-i\gamma K} e^{i\gamma \hat{J_z}}\approx\frac{1}{N_{\gamma}} \sum_{n=1}^{N_{\gamma}}e^{i\frac{2\pi n}{N_{\gamma}}(\hat{J_z}-K)},
\end{equation}    
one can see this is actually the same as the original mesh for our linear algebra inversion on $\alpha, \gamma$, see below. 
This method was also proposed by Fomenko  \cite{Fomenko1970},  as well as independently 
derived as the `Fourier method' in \cite{PhysRevC.49.1422}, used primarily for particle number projection 
\cite{PhysRevC.74.064309,PhysRevC.76.054315,PhysRevC.81.064323}. It is based upon the discrete Fourier identity 
\begin{equation}
\frac{1}{N} \sum_{k=1}^N \exp \left ( i \frac{2\pi M k}{N}  \right ) = \delta_{M,0}. \label{fourier}
\end{equation}
Because Fomenko projection is equivalent to an exact inversion, it makes sense it is superior to Gauss-Legendre quadrature, despite its 
superficial resemblance to trapezoidal quadrature. 

Evaluations can be reduced by using various symmetries, as discussed below. If one has time-reversal symmetry in the original state, it is possible to get a 
$16$-fold reduction \cite{PhysRevC.78.024309,PhysRevC.81.064323} , but for the most general  time-reversal-violating configurations, with odd 
numbers of particles, one can get only a factor of two reduction. 

}
}

\section{Solving linear algebra equations for projection and reduced evaluations}

The central goal of this paper is to reduce the number of evaluations needed for projection by either quadrature 
or linear algebra projection.  In quadrature projection 
this has typically been done by use of symmetries.

As noted above, the central task in projection by linear algebra is solving Eq.~(\ref{compactLA}), where 
 the matrix $D_{ia}$, \textit{must be invertible (nonsingular)}.  Satisfying this condition is not automatic. 

In principle the most efficient method would be to choose the number of Euler angles to 
be the same as the number of angular momentum quantum numbers. Because finding such a minimal set of 
Euler angles which leads to an invertible matrix is difficult, we follow a simpler though somewhat less efficient  
path, where we invert on each Euler angle separately. That is, for the norm we use Eq.~(\ref{WignerD})
and  introduce 
\begin{equation}
n_{ijk} \equiv  \langle \Psi |  \exp( i \alpha_i \hat{J}_z) \exp( i \beta_j \hat{J}_y) \exp( i \gamma_k \hat{J}_z)  | \Psi \rangle \label{nijk} 
\end{equation}
which is equal to
\begin{equation}
 \sum_{JKM} e^{i\alpha_i M} d^J_{MK}(\beta_j) e^{i \gamma_k K}\,N^{J}_{K, M}.
\end{equation}
As proposed  previously \cite{PhysRevC.96.064304}, we first invert on $\alpha, \gamma$, that is, to project out $M, K$, and then on $\beta$ to project $J$.
For $\alpha$, $\gamma$ we originally chose as a mesh 
$ \alpha_i = (i-1) \frac{2\pi}{2J_\mathrm{max} +1}$ for $i=1,\ldots, 2J_\mathrm{max} +1$ and the 
same for $\gamma_k$. For this set of angles, and if $M_a = -J_\mathrm{max}, \ldots, J_\mathrm{max}$
 the square matrix
\begin{equation}
\zeta_{i,a} = \exp \left( i \alpha_i M_a \right )
\end{equation}
can be inverted analytically to get $ \mathbf{Z}=\mathbf{\zeta}^{-1} $, and then obtain the intermediate result
\begin{equation}
n_{j,MK} = \sum_{ik} Z_{Mi} Z_{Kk} n_{ijk} = \sum_J d^J_{MK}(\beta_j) N^J_{MK}. \label{jbetatransform}
\end{equation}
{{
As discusssed in section \ref{quadrature}, this is formally the same as Fomenko projection \cite{Fomenko1970} used to project out 
good particle numbers, and, while arrived at differently, formally the same as trapezoidal quadrature frequently used on $\alpha, \gamma$ 
\cite{hara1995projected,PhysRevC.78.024309}.
}}

Now one needs to invert on $\beta$ to get $J$. The matrix 
$A_{j,a} = d^{J_a}_{MK}(\beta_j)$ (which implicitly depends upon $M,K$) is generally non-square. 
We instead construct the square matrix 
\begin{equation}
\Delta^{J^\prime J}_{MK} = \sum_j d^{J^\prime}_{MK}(\beta_j) d^{J}_{MK}(\beta_j), \label{DeltaJJdefn}
\end{equation}
with $J, J^\prime \geq |M|, |K|$.  This square matrix must be invertible for all the required values of $M, K$. 
{{
Note that all these matrices to be inverted are small. If the maximum $J$ is 20, then these matrices are of dimension $41 \times 41$, and inversion 
takes a tiny fraction of time compared to evaluation of the kernels.
}}

Now we would like to reduce the number of evaluations. We do this in two ways. The first is to 
use  symmetries, so that we get some evaluations for free.  This strategy is  widely used 
in projection by quadrature, see e.g. \cite{hara1995projected,sun1997fortran}.  
 The second is more subtle: if we know $f_J$ is zero or very small for some values 
of $J$, we should not need to include that value of $J$ in our inversion, which in turn can lead to a smaller 
set of evaluations, a strategy we call `need-to-know.'  For example, in some cases for even-even nuclides, the time-reversed-even Hartree-Fock state contains only 
even values of $J$; for another example, if one cranks the Hartree-Fock state by adding an external field, typically 
$\hat{J}_z$, only some high values of $J$ are occupied. 
In both cases, however, one has to find a mesh of angles for which the linear algebra problem is
 solvable, i.e., the matrices are invertible. 

\subsection{Reduction by symmetries}

We start with Eq.~(\ref{rotation}) and  use
\begin{equation}
e^{i\beta \hat{J}_y}= e^{-i \pi \hat{J}_z} e^{- i\beta \hat{J}_y}e^{i \pi \hat{J}_z} 
\end{equation}
so that 
\begin{equation}
\hat{R}(\alpha, \beta, \gamma) = \hat{R} (\alpha-\pi, - \beta, \gamma+\pi) = \hat{R} (\alpha-\pi, - \beta, \gamma-\pi) e^{i 2\pi \hat{J}_z} 
\end{equation}
Then we use for a Slater determinant $| \Psi \rangle$ of fixed number of particles $A$, 
\begin{equation}
e^{i 2\pi \hat{J}_z} | \Psi \rangle = (-1)^A | \Psi \rangle.
\end{equation}
This is easy to see. For a state of fixed $M$, $e^{i 2\pi \hat{J}_z} |M \rangle = \exp( i 2\pi M) |M \rangle$.
For $M$ integer, the phase is $+1$, and for $M$ half-integer, the phase is $-1$; these correspond to $A$ being even or odd, 
respectively. 

Putting this all together, we have 
\begin{eqnarray}
\langle \Psi^\prime | \hat{R}(\alpha, \beta, \gamma) | \Psi \rangle 
& = &  (-1)^A \langle \Psi^\prime | \hat{R}(\alpha-\pi , -\beta, \gamma-\pi) | \Psi \rangle  \nonumber \\
& = & (-1)^A \langle \Psi | \hat{R}(\pi -\gamma , \beta, \pi-\alpha) | \Psi^\prime \rangle^*. 
\end{eqnarray}

Now we can apply these relations in 4 cases:

\begin{eqnarray}
\langle \Psi^\prime | \hat{R}(\alpha, \beta, \gamma) | \Psi \rangle  =   \nonumber \\
 (-1)^A \langle \Psi | \hat{R}(\pi -\gamma , \beta, \pi-\alpha) | \Psi^\prime \rangle^*, \,\,\, & 0 < \alpha, \gamma < \pi  \\
   \langle \Psi | \hat{R}(3\pi -\gamma , \beta, \pi-\alpha) | \Psi^\prime \rangle^*, \,\,\, & 0 < \alpha< \pi <  \gamma < 2\pi  \\
 \langle \Psi | \hat{R}(\pi -\gamma , \beta, 3\pi-\alpha) | \Psi^\prime \rangle^*, \,\,\, & 0 < \gamma< \pi <  \alpha < 2\pi \\
 (-1)^A \langle \Psi | \hat{R}(3\pi -\gamma , \beta, 3\pi-\alpha) | \Psi^\prime \rangle^*, \,\,\, &  \pi < \alpha, \gamma < 2\pi 
\end{eqnarray}

The next step is to find an invertible mesh, that is, a set of $2J_\mathrm{max}$ angles $\{ \gamma_k\}$ such that the matrix 
$\zeta_{ka} = \exp(i \gamma_k M_a)$ is numerically invertible, where $M_a = -J_\mathrm{max}, -J_\mathrm{max}+1, \ldots, +J_\mathrm{max}$.  We have found such a mesh:

\medskip

\noindent  Case 1:  $\mathrm{mod} (2J_\mathrm{max}+1,4)=1$, or $J_\mathrm{max}=0,2,4,6,\ldots$.

Let $\nu = J_\mathrm{max}$. Then choose
$$
\gamma_k = \left \{ 
\begin{array}{l}
\pi\frac{k}{\nu+1}, \,\, k= 1,\nu;\\
\pi+\pi\frac{(k-\nu)}{\nu+1}, \,\, k= \nu+1,2\nu \\
\pi ,\,\, k = 2\nu + 1 = 2 J_\mathrm{max} +1
\end{array} \right .
$$

\bigskip

\noindent  Case 2:  $\mathrm{mod} (2J_\mathrm{max}+1,4)=2$, or $J_\mathrm{max}=1/2,5/2,9/2,\ldots$.

Let $\nu = J_\mathrm{max}-\frac{1}{2}$. Then choose 
$$
\gamma_k = \left \{ 
\begin{array}{l}
\pi\frac{k}{\nu+1}, \,\, k= 1,\nu;\\
\pi+\pi\frac{(k-\nu)}{\nu+1}, \,\, k= \nu+1,2\nu \\
0 ,\,\, k = 2\nu + 1 = 2 J_\mathrm{max} \\
\pi , \,\,\, k = 2\nu + 2 = 2 J_\mathrm{max} +1
\end{array} \right .
$$

\bigskip

\noindent  Case 3:  $\mathrm{mod} (2J_\mathrm{max}+1,4)=3$, or $J_\mathrm{max}=1,3,5,7,\ldots$.

Let $\nu = J_\mathrm{max}-1$. Then choose 
$$
\gamma_k = \left \{ 
\begin{array}{l}
\pi\frac{k}{\nu+1}, \,\, k= 1,\nu;\\
\pi+\pi\frac{(k-\nu)}{\nu+1}, \,\, k= \nu+1,2\nu \\
0 ,\,\, k = 2\nu + 1  \\
\pi/2, \,\, k =  2\nu + 2  = 2 J_\mathrm{max}\\
\pi , \,\,\, k = 2\nu + 3  = 2 J_\mathrm{max} +1
\end{array} \right .
$$

\bigskip

\noindent  Case 4:  $\mathrm{mod} (2J_\mathrm{max}+1,4)=0$, or $J_\mathrm{max}=3/2,7/2,11/2,\ldots$.

Let $\nu = J_\mathrm{max}+\frac{1}{2}$. Then choose 
$$
\gamma_k = \left \{ 
\begin{array}{l}
\pi\frac{k}{\nu+1}, \,\, k= 1,\nu;\\
\pi+\pi\frac{(k-\nu)}{\nu+1}, \,\, k= \nu+1,2\nu 
\end{array} \right .
$$

\medskip

With this mesh, one can get $\mathbf{Z} = \mathbf{\zeta}^{-1}$, albeit numerically.
{{
The dimensions are small so inversion is quick.
}}

In principle one could use additional symmetries that include $\beta$. Those symmetries, however, 
generally require some sort of axial symmetry and work only for even-even nuclides. Because we often 
work with odd-$A$ or odd-odd nuclides, and our Hartree-Fock code \cite{PhysRevC.66.034301} allows for general triaxiality, 
we did not pursue additional symmetries. 
{{
We numerically confirmed the above mesh is invertible for all $J_\mathrm{max}  \leq 30$, including half-integers.
}}
We implemented this 
{{symmetry}}
 and confirmed its accuracy and speed-up.
In our results in Section \ref{results}, however, we did not use this {{or any}} symmetry.

\subsection{Reduction by need-to-know}

The basic idea of projection by linear algebra is to solve Eq.~(\ref{compactLA}) and related equations, which yields the norm 
and Hamiltonian matrices in Eq.~(\ref{eigen}).  If, however, $f_J \approx 0$, then there is no need to solve for the matrices for that 
$J$.   Thus we  can consider reducing the number of $J$-values in $\Delta^{J^\prime,J}$, which in turn allows 
one to reduce the mesh on $\beta$.  

Previously we found empirically a simple invertible mesh \textit{if} one includes all $J \leq J_\mathrm{max}$:
\begin{equation}
\beta_j = (j-1/2) \frac{\pi}{N}, \,\,\ j = 1,N\label{betamesh}
\end{equation}
where $N= J_\mathrm{max}+1$ if an even system and $=J_\mathrm{max}+1/2$ if an odd number of nucleons.
By eliminating some values of $J$ one should be able to also reduce the number of points on $\beta$ one evaluates. 
This turned out to be nontrivial: simply eliminating some of the $\beta_j$, or rescaling, led to singular or near-singular 
matrices $\Delta^{J^\prime,J}$. 

With some experimentation, however, we found a procedure which for 
most cases yielded  a mesh which led to invertible $\Delta^{J^\prime,J}$ for all the values of $K,M$. 
The criterion for invertibility is that the eigenvalues of $\Delta^{J^\prime J}_{MK}$
are nonzero for all desired values of $M,K$ (remember we only treat $J^\prime, J$ as the indices of the matrix). 
We started with (\ref{betamesh}), except that $N = $ the number of ``occupied'' $J$ values (given by some 
critical value of $f_J$). We then looped over all $M,K$ and found the eigenvalues of $\Delta^{J^\prime J}_{MK}$; 
because the dimensions are small this is extremely fast. We then counted how many eigenvalues were below a 
threshhold $\epsilon$, and also computed the sum of these near-singular eigenvalues.  We then swept through the $\beta_j$, randomly perturbing their values. If the number of 
near-singular eigenvalues decreases, or the sum of near-singular eigenvalues increased (without increasing the 
number of near-singular eigenvalues), the change in $\beta_j$ is accepted.  

While this process can take several dozen 
sweeps, the overall time burden is small. In some cases, however, our simple Monte Carlo procedure failed to 
find invertible solutions. To date we do not have a theory as to when solutions can and cannot be found.  We also emphasize 
that if we do not insist on need-to-know and simply use a $J_\mathrm{max}$, our meshes have always been invertible.


\section{Results}

\label{results}
\begin{figure}
\centering
\includegraphics[scale=0.45,clip]{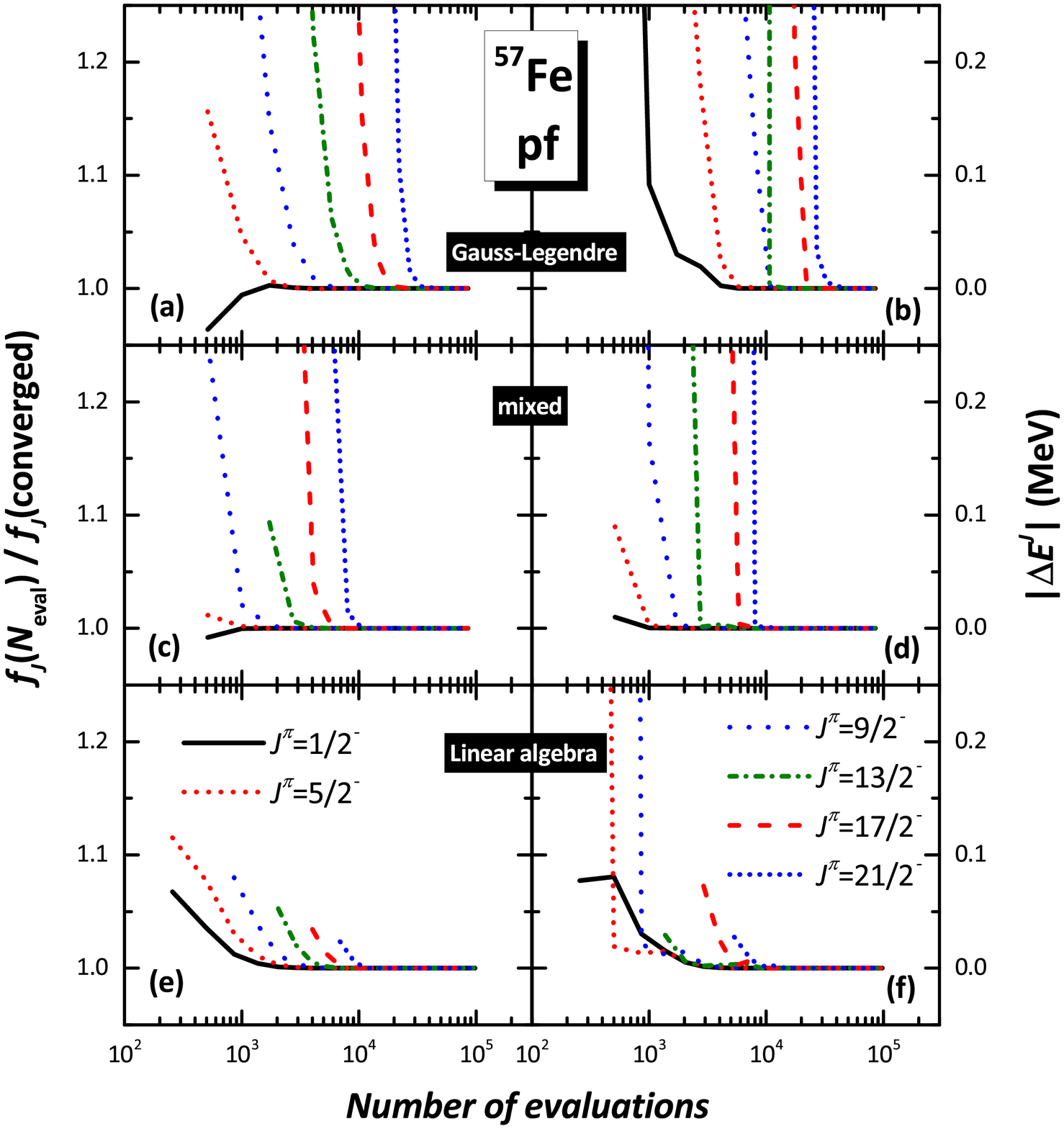}
\caption{(Color online) 
Convergence, as a function of the number of evaluations,  of projection via 
{{
Gauss-Legendre}}
quadrature (upper panels), 
{{
via mixed quadrature, that is,  inversion in 
the $\alpha,\gamma$ angles, equivalent to Fomenko projection/trapezoidal quadrature, as discussed in Section \ref{quadrature},
and Gauss-Legendre quadrature on $\beta$ (middle panels),
}}
and via linear algebra (lower panels)
of the fractional occupation $f_J$ (left panels) and the yrast energies (right panels) 
for $^{57}$Fe in the $0f$-$1p$ space. }
\label{Fefig}
\end{figure}

We remind the reader that our criterion for efficiency is the \textit{number of evaluations} at different Euler angles required for a 
converged result (i.e., does not change with increased number of evaluations), and that evaluation of the norm (overlap) kernel is computationally much cheaper than for the Hamiltonian kernel. 
With that in mind, we broadly found that projection by linear algebra requires significantly fewer evaluations than quadrature. 
Furthermore, we found that convergence of the norm kernel, represented by the fractional occupation $f_J$, tracks the 
convergence of the Hamiltonian kernel and the resulting energies. 

This we illustrate with three nuclides in different model spaces with different semi-phenomenological interactions
in Figs.~\ref{Fefig}, \ref{Gafig}, and \ref{Crfig}, showing the convergence of $f_J$ and the yrast energies as a function of the 
number of evaluations, $N_\mathrm{eval}$.  Specifically, we show the ratio $f_J(N_\mathrm{eval})/f_J(\mathrm{converged})$ 
in the left-hand panels, and the difference in the yrast energies $E^J(N_\mathrm{eval}) - E^J(\mathrm{converged})$ in the 
right-hand panels.  
{{
In Fig. \ref{Crfig} we also show the convergence of $J^2$, which we obtained by replacing the Hamiltonian matrix elements with those of 
$J^2$.}}



\label{results}
\begin{figure}
\centering
\includegraphics[scale=0.4,clip]{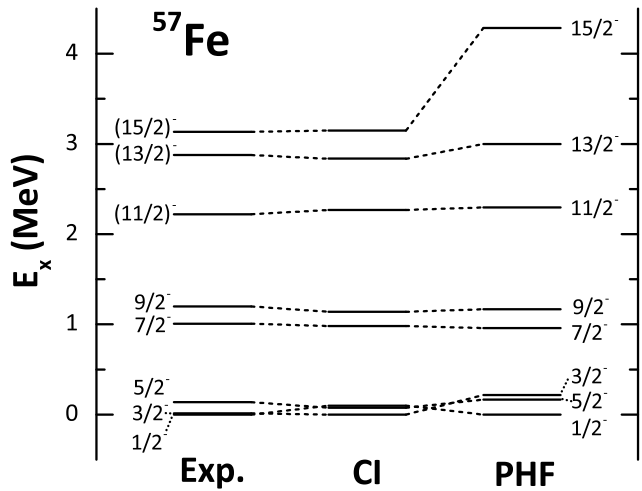}
\caption{(Color online) 
{{
Low-lying levels of 
 $^{57}$Fe in the $0f$-$1p$ space, comparing experimental and full configuration-interaction (CI) calculations with results from projected Hartree-Fock (PHF). }
 }}
\label{Fespect}
\end{figure}

Our specific examples are: Fig.~\ref{Fefig}, $^{57}$Fe in the $0f$-$1p$ shell with frozen $^{40}$Ca core and the monopole-modified $G$-matrix interaction GX1A \cite{honma2005shell}; Fig.~\ref{Gafig}, $^{68}$Ga
in the $0f_{5/2}$-$1p$-$0g_{9/2}$ space with a frozen $^{56}$Ni core, and the interaction JUN45 \cite{PhysRevC.80.064323};
and finally Fig.~\ref{Crfig}, $^{48}$Cr in the $1s$-$0d$-$0f$-$1p$ shells with frozen $^{16}$O core, 
with the interaction of \cite{PhysRevLett.116.112502}.  
Results for other nuclides are similar and not sensitive to the model space; for example, results for $^{48}$Cr 
in $0f$-$1p$ shell are qualitatively indistinguishable from Fig.~\ref{Crfig}.  Calculations with other nuclides, in these spaces and others, behave in very similar fashion. This includes preliminary results in even larger, multi-shell spaces. 

{{
We also show, in Fig.~\ref{Fespect}, the low-lying levels of $^{57}$Fe from experiment \cite{nndc}, from full configuration-interaction shell model calculations, and 
using projected Hartree-Fock, with the latter two both using GX1A in the $pf$ space.  In general we find projected Hartree-Fock does well in reproducing spectra of 
even-even nuclei, especially those which are strong rotors, poorly for odd-odd nuclei, and mixed quality for odd-$A$ nuclei. $^{57}$Fe happens to be reasonably well-reproduced, 
and we speculate this may be due to being a simple particle plus rotor. In general one would want to go to configuration-mixing via, for example, the generator-coordinate method 
\cite{PhysRevC.65.024304,PhysRevC.74.064309,PhysRevC.78.024309,PhysRevC.81.044311,PhysRevC.81.064323,PhysRevC.83.014308,borrajo2015symmetry,Rodríguez2016}. Our purpose here is primarily to investigate the convergence of the projection method.
}}

\begin{figure}
\centering
\includegraphics[scale=0.45,clip]{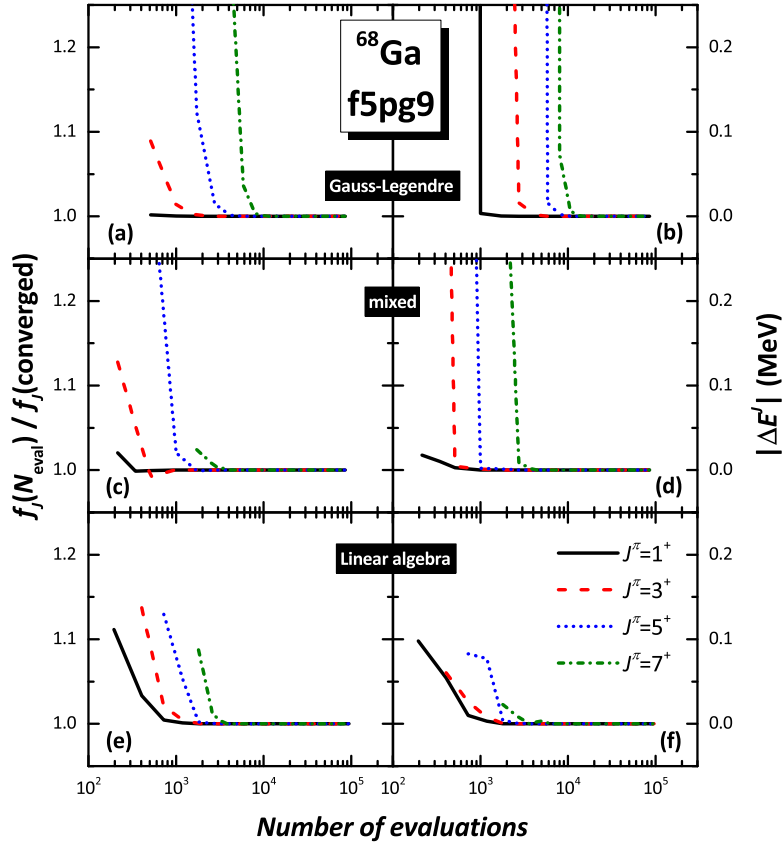}
\caption{(Color online) Convergence, as a function of the number of evaluations, of projection via Gauss-Legendre quadrature (upper panels),
{{
via mixed quadrature, that is,  inversion in 
the $\alpha,\gamma$ angles, equivalent to Fomenko projection/trapezoidal quadrature, as discussed in Section \ref{quadrature},
and Gauss-Legendre quadrature on $\beta$ (middle panels),
}}
and via linear algebra (lower panels) 
of the fractional occupation $f_J$ (left panels) and the yrast energies (right panels) 
for $^{68}$Ga in the $0f_{5/2}$-$1p$-$0g_{9/2}$ space.
.}
\label{Gafig}
\end{figure}

{{
In Figs.~\ref{Fefig}, \ref{Gafig}, and \ref{Crfig}, we show three projection methods. 
}}
For projection by Gauss-Legendre quadrature, we assumed the same number of mesh points $N_\Omega$ for all three Euler angles,  so that the number of evaluations is $N_\Omega^3$.  
We found $N_\Omega = 44$ (or $N_\mathrm{eval} = 85,184$) produced reliably converged results.
 Smaller values of $J$ converge faster with $N_\Omega$ than larger values, which 
makes sense: one expects the large $J$ wave functions to have more nodes in $\alpha, \beta, \gamma$. 
For projection by  linear algebra, we increased $J_\mathrm{max}$ until we got no change in results;   increasing 
$J_\mathrm{max}$ further made no difference. The number of evaluations is roughly $4 J_\mathrm{max}^3 + 8 J_\mathrm{max}^2$.
{{
We also show a `mixed' projection, where the $\beta$-integral, for rotations about the $y$-axis, is done by Gauss-Legendre quadrature,
while the $\alpha$- and $\gamma$-integrals, for rotations about the $z$ axis, is done by linear algebra inversion; as discussed 
in section \ref{quadrature} this is equivalent to so-called `trapezoidal' quadrature in the literature as well as Fomenko projection 
 \cite{Fomenko1970}.}} In these 
results we did not use symmetries to reduce the number of evaluations in either method.

{{
In Figs.~\ref{Fefig} and \ref{Gafig}, we oriented the plots so that, reading downward, one can see the improved convergence as
one goes from full Gauss-Legendre quadrature, to mixed projection, to linear algebra inversion.
For Fig.~\ref{Crfig}, we changed the orientation, so that, reading down, one can compare the convergence of the 
fractional occupation $f_J$, the yrast energies, and of $J^2$.  The latter we treated by simply replacing the Hamiltonian with the 
matrix elements for $J^2$.  One sees that all three quantities converge at similar points, although the yrast energies converge 
with a slightly larger number of evaluations.  All our other results, not only for $^{57}$Fe and $^{68}$Ga but many other nuclides, 
exhibit qualitatively the same behavior. 

}}


\begin{figure}
\centering
\includegraphics[scale=0.28,clip]{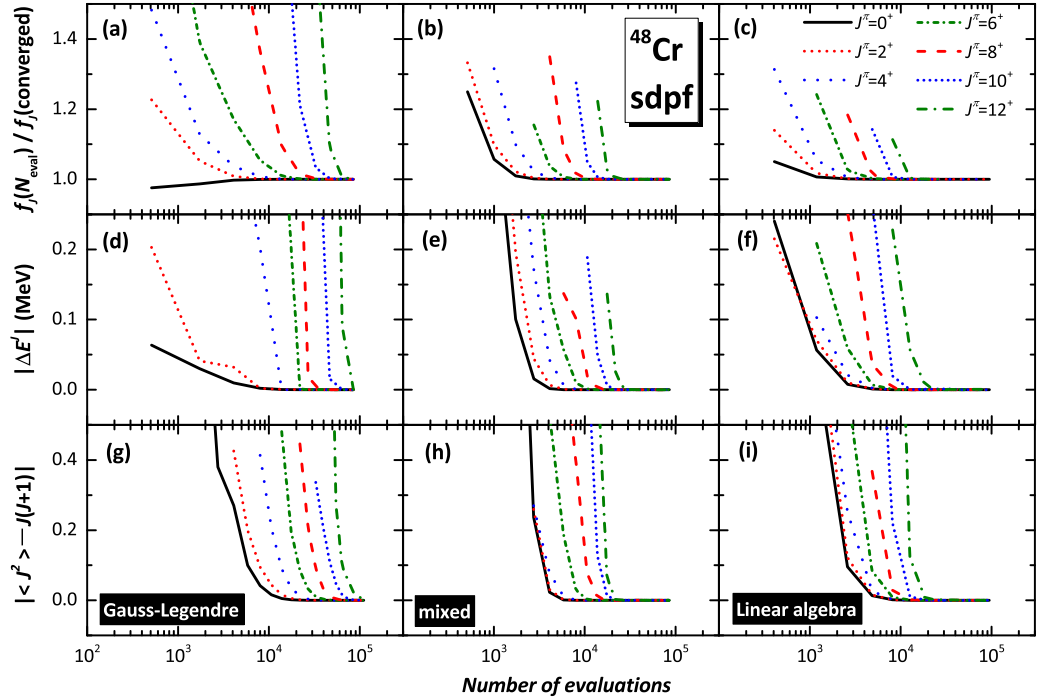}
\caption{(Color online) 
{{
Note: orientation and content differ from Figs.~\ref{Fefig},\ref{Gafig}. Convergence, as a function of the number of evaluations,  of projection via Gauss-Legendre  quadrature (left-hand panels),
via mixed quadrature, that is,  inversion in 
the $\alpha,\gamma$ angles, equivalent to Fomenko discretization/trapezoidal quadrature, as discussed in Section \ref{quadrature},
and Gauss-Legendre quadrature on $\beta$ (middle column panels),
and via linear algebra (right-hand panels)
of the fractional occupation $f_J$ (top row panels), yrast energies (middle row panels), and expectation values of $J^2$ (bottom 
row panels)
for $^{48}$Cr in the $0d$-$1s$-$0f$-$1p$ space.
}
}}
\label{Crfig}
\end{figure}


Our results are summarized in Table \ref{evaluations}, which shows the number of evaluations needed to get specific energies 
to within 1 keV of the converged results.   This table clearly shows the advantage of projection by linear algebra, 
{{
which includes
for us the `mixed' case of trapezoidal `quadrature'/Fomenko projection.}}
 For low $J$ 
the advantage is small, largely because in projection by quadrature one can target a specific value of $J$, say $J=0$, 
while in projection by linear algebra one needs to solve for all, or at least the most important (as measured by $f_J$) 
values of $J$. For larger $J$, however, projection by linear algebra robustly requires a factor of 3 fewer evaluations, or better, 
{{than pure Gauss-Legendre quadrature.}}
Note that $J$ in Table \ref{evaluations} is not  $J_\mathrm{max}$; $J_\mathrm{max}$ is determined by the criterion 
of convergence to within 1 keV.

{{
For comparison, consider some typical number of evaluations in the literature.
For projection up to $J=6$, we found reported meshes of 20,000 evaluations for mixed projection \cite{PhysRevC.78.024309}, 
and for pure Gauss-Legendre quadrature 18,000 \cite{PhysRevC.81.064323} to 32,000 \cite{PhysRevC.79.044312}.
In a case for up to only $J=2$, Gauss-Legendre only required 3,500 evaluations \cite{PhysRevC.79.044312}. 
It has also been reported that larger deformations, mixing in higher $J$-levels, require significantly more evaluations 
\cite{PhysRevC.81.064323}.
Note: all these calculations considered triaxially deformed but time-reversed-even states, and so used symmetries 
to reduce the number of actual evaluations; we multiplied by 16 in order to compare.
These numbers are not too different from our values in Table 1, although we found our mixed calculations required 
a factor of 2 fewer evaluations to get to $J=6$. 
 }}

We also experimented with our implemented  need-to-know algorithm, which, by knowing that some values of $f_J \approx 0$, 
we can reduce the number of values of $\beta_j$ at which we need to evaluation. In some cases we could reduce the number 
of evaluations by a factor of two, but we found these to be rather specialized cases. 

To be specific: for some even-even nuclides, the Slater determinant is time-reversal even, and only even values of $J$ are 
occupied (have non-zero $f_J$). In the $0f$-$1p$ shell, for example, both $^{48}$Cr and $^{60}$Fe are prolate, and we were 
able to reduce the number of evaluations from 12,615 to 6,728
with a change in energies by less than 0.4 keV. 
 $^{62}$Ni  is oblate; using need-to-know we reduced the number of evaluations from 18,513 to 9,801, 
with a change in energies less than 0.6 keV.  In all these cases our Monte Carlo algorithm quickly found a mesh of $\beta_j$ 
which was invertible. 

We also tried need-to-know with cranked wave functions: by adding $-\omega \hat{J}_x$ (or any other component of angular 
momentum) the solution Slater determinant contains higher fractions of components at higher $J$, and smaller fractions at smaller $J$.  The saving in evaluations, however, is generally small,
except for large values of $\omega$. In contrast to the time-reversed even cases, with only even values of $J$, our empirical 
experience is that it is more difficult to find invertible meshes when deleting small values of $J$ and keeping large values; the 
reason remains unclear to us.

\begin{table}[h]
\begin{center}
\caption {
{{
The minimum number of evaluations needed for $^{57}$Fe, $^{68}$Ga, and $^{48}$Cr when $|\Delta E^J| = |E^J(N_{\mathrm{eval}})-E^J(\mathrm{converged})|$ to be smaller than 1 keV. 
Here `Gauss-Legendre' is Gaussian quadrature in all three Euler angles, `lin.~alg.' inverts in all three Euler angles, and `mixed' means inversion in 
the $\alpha,\gamma$ angles, equivalent to Fomenko discretization/trapezoidal quadrature, as discussed in Section \ref{quadrature}, with Gauss-Legendre quadrature in the $\beta$ angle. 
}}
\label{evaluations}}
\begin{tabular}{m{1cm}<{\centering} lcccc}
\hline
\multirow{5}{*}{Nuclide} &\multirow{5}{*}{$J^{\pi}$} & \multicolumn{3}{c}{\multirow{1}{*}{$N_{\mathrm{eval}}$}}   \\
& & \multicolumn{3}{c}{\multirow{1}{*}{($|\Delta E^J| < 1$ keV)}}\\
& & \cline{1-3} && {Gauss-Legendre}  & {mixed} & {lin. alg.} \\
\hline
$^{57}$Fe & $1/2^+$ & 5832 &1000& 4000 \\
                 & $5/2^+$ &8000 &1728 &4000\\
                 & $9/2^+$ &13824 &2744& 5324 \\
                 & $13/2^+$ & 13824&5832& 5324 \\
                 & $17/2^+$ &27000 & 8000&8788 \\
                 & $21/2^+$ & 32768 &10648 &13500 \\
\hline
$^{68}$Ga & $1^+$ & 1728 &1000&1800 \\
                  & $3^+$ &5832 &1728 &1800 \\
                  & $5^+$ & 10648 &2744& 2601\\
                  & $7^+$ &13824 &4096 &3610 \\
\hline
$^{48}$Cr  & $0^+$ & 13824&5832 &8125 \\
                  & $2^+$ &13824 &5832 &8125\\
                  & $4^+$ & 13824&8000 &8125\\
                  & $6^+$ & 21952&10648&8125 \\  
                  & $8^+$ &46656 &17576 &12615\\ 
                  & $10^+$ & 64000& 21952&12615\\  
                  & $12^+$ &85184 &32768 &18513\\                      
\hline
\end{tabular}
\end{center}
\end{table}

\section{Conclusions and acknowledgements.}

We have investigated two related  approaches for projection of good angular momentum, 
projection by quadrature and projection by linear algebra. In both methods one samples matrix 
elements on a mesh of Euler angles; because evaluation of Hamiltonian matrix elements,
is computationally expensive we want to use  a minimal mesh. In particular we investigated
the convergence of $f_J$, the fraction of the wave function 
with angular momentum $J$, because the sum of $f_J$ must be 1, and because it is cheaper to compute overlaps than energies. 
Therefore $f_J$ is our suggested key criterion for convergence, for both methods.

{{
The situation is somewhat complicated by the fact that many papers cite a trapezoidal `quadrature' method. 
Although justification for this method is hard to find in the literature, in our analysis this method is equivalent to our linear 
algebra inversion, as well as to the Fomenko projection cited in particle-number projection based upon an exact finite Fourier sum rule 
\cite{Fomenko1970,PhysRevC.49.1422}.  In all cases, however, using linear algebra inversion methods, including 
trapezoidal `quadrature' / Fomenko projection, dramatically reduce the number of evaluations required;} }
If one is interested in projecting 
out states with all or most values of $J$, the reduction is as much as three-fold, i.e., only as third as many evaluations are required
{over Gauss-Legendre  quadrature.}

In all methods one can reduce the number of evaluations by using discrete symmetries. Because we do not impose 
axial symmetry upon our Hartree-Fock solutions, we only considered symmetries in 
the Euler angles $\alpha, \gamma$ (rotations about the $z$-axis). We found meshes 
which met the symmetry but still lead to invertible matrices.  If one imposed axial symmetry, there are additional 
possible savings, which we did not explore.

In some cases it is possible to get additional gains 
by eliminating unoccupied values of $J$ and to reduce simultaneously the number of evaluations on 
the Euler angle $\beta$ (rotation about the 
$y$-axis). Aside from time-reversed-even cases, however, the gains were generally not large.

This material is based upon work supported by the U.S. Department of Energy, Office of Science, Office of Nuclear Physics, 
under Award Number  DE-FG02-03ER41272.   {{
We are grateful to the anonymous referee who nudged us towards Fomenko projection, 
which led us a better understanding of the literature.}}

\bigskip

\bibliographystyle{unsrt}
\bibliography{johnsonmaster}

\end{document}